\documentclass[aps,prl,twocolumn,superscriptaddress,longbibliography]{revtex4-2}
\usepackage{amsmath,amssymb,bm,graphicx}

\usepackage{dcolumn}

\usepackage{accents}

\usepackage{hyperref}

\usepackage{xcolor}

\usepackage{ulem}

\usepackage{comment}

\begin{document}

\title{Experimentally probing the Quantum Physics in the Inverted Harmonic Oscillator}

\author{Si-Cong Ji}
\author{Philipp Schüttelkopf}
\author{Nataliia Bazhan}
\author{Federica Cataldini}
\author{Mohammadamin Tajik}
\author{Frederik S. Møller}
\author{Igor Mazets}
\author{Sebastian Erne}
\author{Jörg Schmiedmayer}
\affiliation{Vienna Center for Quantum Science and Technology (VCQ), Atominstitut, TU Wien, Vienna, Austria}

\date{\today}

\begin{abstract}
When a quantum system passes through an unstable fixed point the local dynamics reduces to the inverted harmonic oscillator (IHO). It exponentially amplifies along one quadrature while squeezing the other, producing macroscopically extended quantum states from microscopic zero-point fluctuations. We realize this dynamics with a Bose--Einstein condensate on an AtomChip. Radio-frequency dressing flips the transverse harmonic confinement into an IHO. Through phase-space tomography we follow the full Wigner function of the evolving quantum state, observe sub-vacuum squeezing of $10.6(1.3)$~dB, and test coherent reversibility by time-reversing the IHO evolution.  Matter-wave interference between the two daughter clouds confirms quantum coherence over timescales far beyond the initial expansion. 
Our experiment establishes ultra-cold atoms as a clean, controlled, many-body platform for unstable quantum dynamics opening a route to force sensing with time-reversal-based coherence certification and to analog studies of the amplification of quantum fluctuations in inflationary field dynamics.
\end{abstract}

\maketitle

%\section{Introduction}
\noindent\textbf{Introduction }\\
The inverted harmonic oscillator (IHO) is the paradigmatic model of unstable quantum dynamics.  With its hyperbolic phase-space flow, it generates exponential stretching of one quadrature and simultaneous squeezing of its conjugate.  This %universality as the local quantum description of unstable fixed points 
endows the IHO with a role across physics that ranges far beyond a mathematical curiosity: from squeezed light in quantum optics~\cite{Barton1986,Wu1986,Lvovsky2015}, saddle-point instabilities in the lowest-Landau-level and quantum Hall systems~\cite{Subramanyan2021,Hegde2019,Sierra2008}, to the local dynamics near black-hole horizons and the onset of chaos~\cite{Franz2018,Hashimoto2017,Dalui2019,Dalui2020}. 

Most directly relevant to the present work, the IHO is mathematically identical to the dynamics of inflationary perturbations,  where vacuum fluctuations are stretched into macroscopic, squeezed quantum states~\cite{Albrecht1994}. This parallel, exact at the level of mode equations,  motivates the term \emph{coherent inflation} (CI) introduced by Romero-Isart and co-workers~\cite{Rom17,Pin18,Wei21}: by preparing a motional ground state at the unstable fixed point of an IHO, fragile zero-point fluctuations are amplified into macroscopic quantum delocalizations. Combined with the \emph{loop protocol}~\cite{Wei21}, in which the IHO evolution is time-reversed to recompress the state, this provides both a route to large spatial superpositions of massive objects and a means of certifying coherence.  Applications range from matter-wave interferometry and force sensing, to tests of gravitationally induced decoherence~\cite{Bel18,Geraci2015,Hebestreit2018}.

Realizing IHO dynamics experimentally with massive particles requires more than producing a parabolic barrier: it requires that the \emph{quantum} content of the state, that is its zero-point fluctuations, survive the inversion. The originally envisioned platforms of levitated nanoparticles~\cite{Delic2020,Magrini2021,Tebbenjohanns2021,Weiss2019} have recently made important progress in this direction. Tomassi \emph{et al.}~\cite{Tomassi2026} released the center-of-mass thermal state ($\bar{n} \approx 10$ phonons) of a silica nanoparticle into a dark electrostatic inverted potential and observed a $\sim$\,950-fold expansion. The observed growth is the phase-space stretching that a Gaussian distribution of classical particles would exhibit. The distinctively quantum content of CI, the amplification of \emph{zero-point} fluctuations paired with the conjugate quadrature squeezed \emph{below} the vacuum, requires a near-pure initial state, and certification of coherent unitary evolution.

In the present work we realize these conditions in a Bose--Einstein condensate on an AtomChip \cite{Folman2000}. By  fast switching between harmonic and inverted-harmonic transverse confinements and following the evolution by reconstructing the full Wigner function through phase-space tomography we observe the signatures of coherent IHO dynamics through interference, sub-vacuum squeezing of up to $10.6(1.3)$~dB and refocusing.

\begin{figure}
\includegraphics[width=0.95\columnwidth]{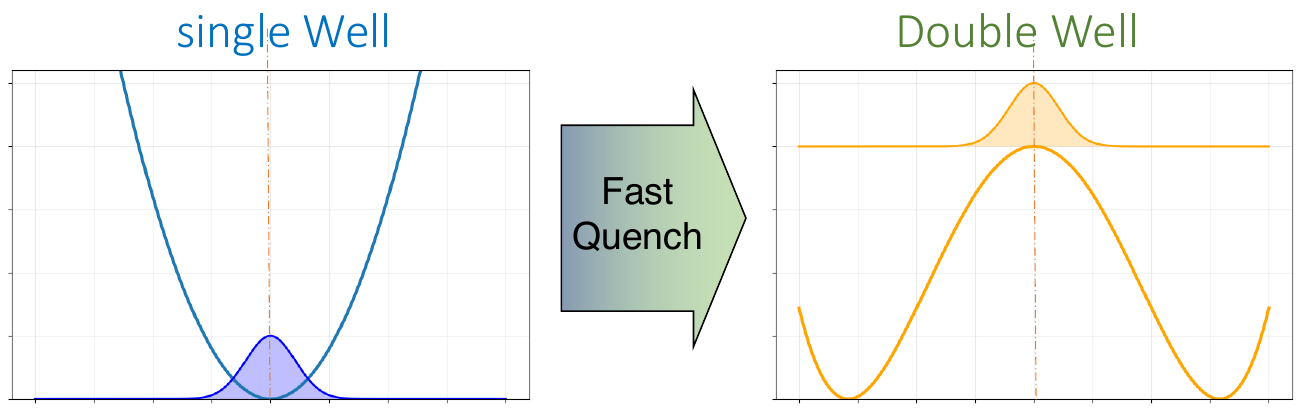}
\caption{\label{fig:SW-Q-DW} 
Preparation of the IHO initial condition. The RF quench transforms the transverse harmonic confinement into a double-well potential whose central barrier is locally approximated by an inverted harmonic oscillator. The atoms are initially localized at the former trap minimum, which coincides with the unstable fixed point of the IHO.}
\end{figure}

\vspace{5mm}\noindent\textbf{Experimental setup } \\
We start the experiment by preparing a quantum degenerate Bose gas of $N\approx 10^4$ $^{87}$Rb atoms in an elongated one-dimensional trap on an AtomChip \cite{Folman2000} with confinement of $\omega_\bot=2\pi \times 2.1$ kHz transversely (radially) and $\omega_l=2\pi \times 10$ Hz longitudinally. At a linear density of $n_{1D}<100$ atoms/µm the interaction energy µ $<\hbar\omega_\bot$ and at ultra low temperatures in the range 20–40 nK ($k_B$T $<$ 0.3 $\hbar\omega_\bot$) the atoms are prepared in the interaction-modified ground state of the radial direction (Fig.~\ref{fig:SW-Q-DW} (left)) with negligible transverse excitations \cite{Kruger2010}.

\begin{figure}
\includegraphics[scale=0.2]{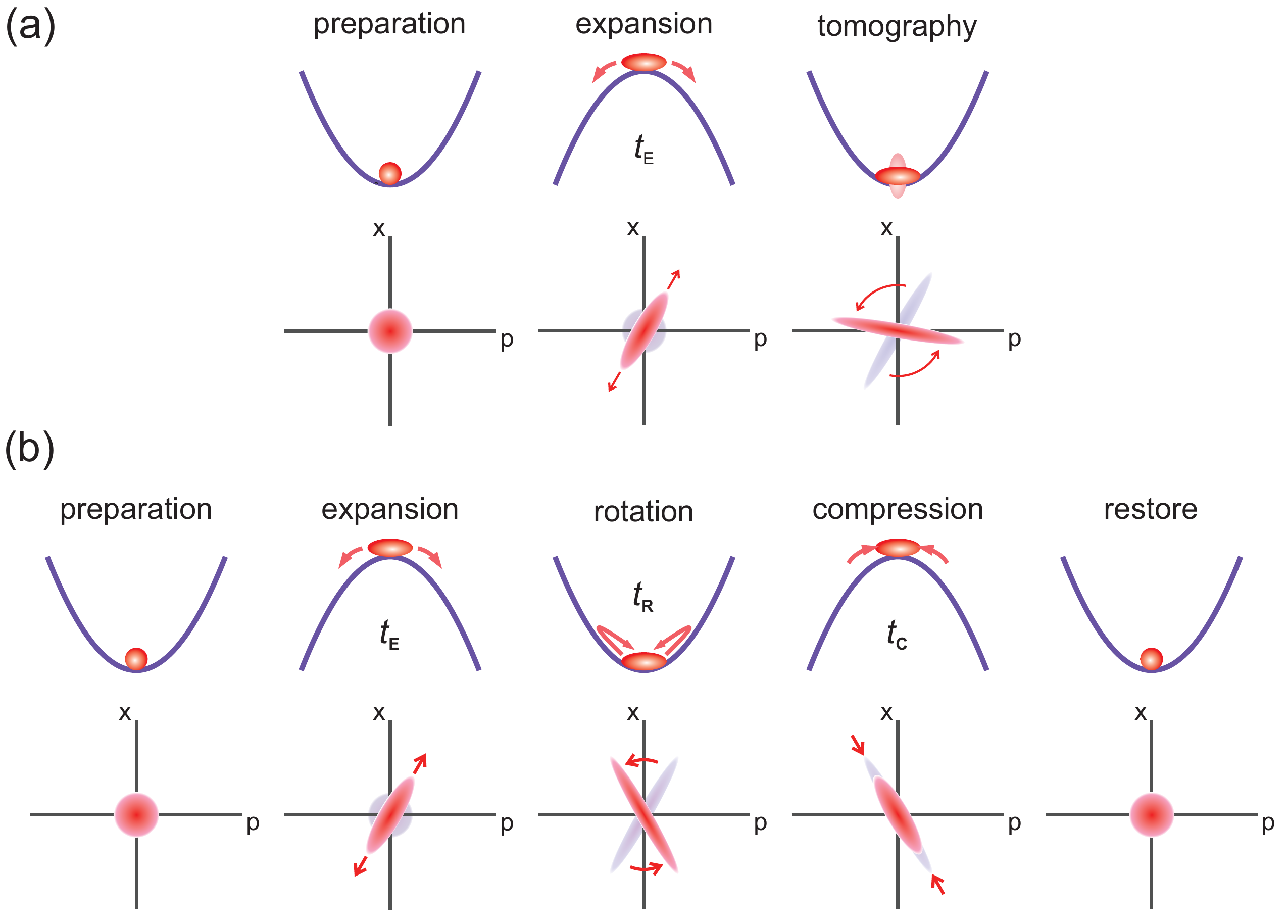}
\caption{\label{fig:Exp_protocal} 
(a) Time sequence of the coherent expansion. (b) Time sequence for restoring the initial coherent state. For both figures, the upper panel shows the potential at different stages of the experiment, while the lower panel shows the corresponding phase-space distributions. The dark (light) red ellipses represent the Wigner distribution at the end (beginning) of each stage.}
\end{figure}

Employing RF dressed state potentials \cite{Hof06,Lesanovsky2006,Lesanovsky2006a}, we quench from the initial harmonic confinement into a double-well potential. Owing to the precise design and fabrication of the AtomChip, the barrier of the double-well is positioned \textit{exactly} at the minimum of the initial harmonic confinement (Fig.~\ref{fig:SW-Q-DW}). The switching time of $\sim 1 \mu s$ is two orders of magnitude faster than the characteristic timescale of transverse motion ($1/\omega_\bot \sim$ 100 µs), the atoms hold their original distribution in both position and momentum, and start their evolution on the top of the potential barrier essentially unperturbed.

Numerical calculations of the RF dressed state potential reveal that within a $1\,\mu$m region surrounding the local maximum between the wells, the inter-well barrier can be effectively treated as an inverted harmonic potential characterized by $\omega_I = 2\pi \times 0.9$ kHz.
By manipulating the RF currents, we can switch the system rapidly between different potential configurations.  This allows us to implement the CI rapid expansion, conduct tomography of the expanding wave function (Fig.~\ref{fig:Exp_protocal} (a)) as well as to implement the loop protocol certifying reversibility by time-reversing the IHO dynamics. (Fig.~\ref{fig:Exp_protocal} (b)).

\begin{figure}
\includegraphics[width=\columnwidth]{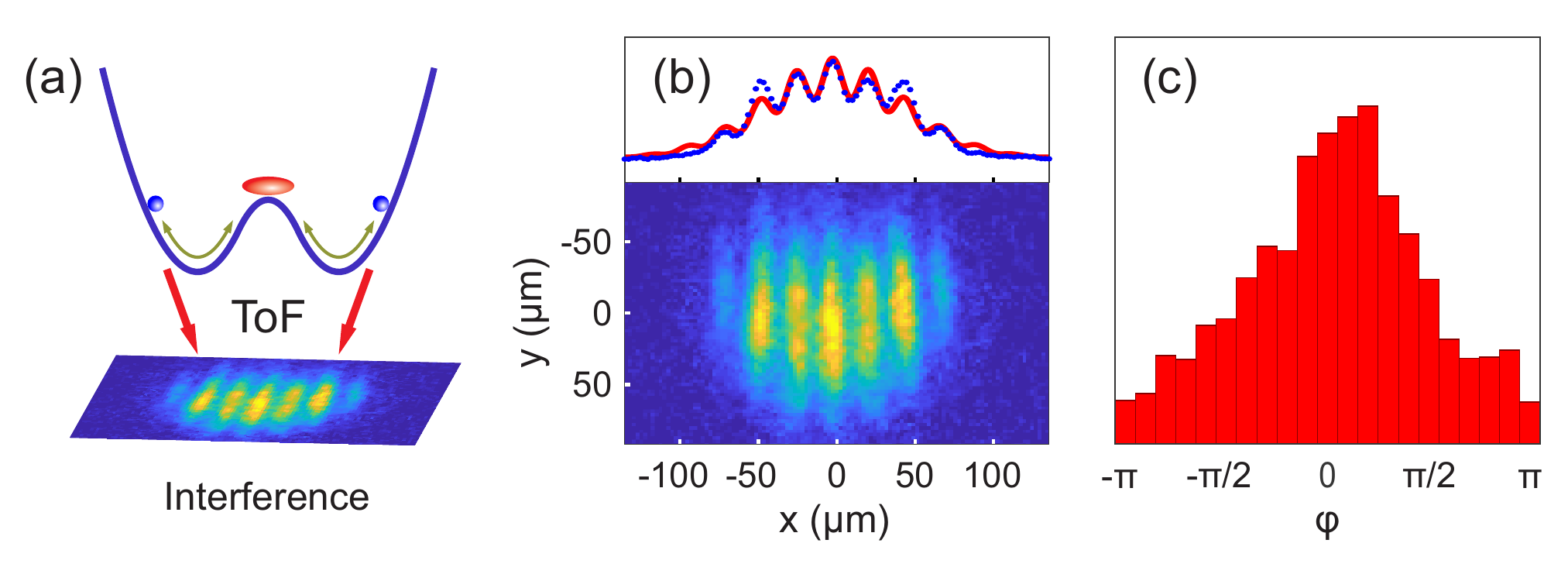}
\caption{\label{fig:Interference}
Coherent expansion probed by matter-wave interference.
(a) The BEC expands in the inverted harmonic potential from the top of the barrier into the double well. The split cloud is released close to the far turning point. The wave functions expand and overlap in time-of-flight (ToF).
(b) Interference pattern of the two daughter clouds split by the barrier after a $5.5$~ms hold, integrated along the longitudinal (z) direction.
(c) Distribution of the  phase $\varphi$ extracted from the fringes. The fringe contrast $C = 0.24(0.04)$, together with the pronounced peak in  $\varphi$ rather than a uniform distribution, demonstrate that phase coherence between the two copies survives the full inflation.
}
\end{figure}

\vspace{5mm}\noindent\textbf{Coherent expansion} \\
We start with studying the coherent expansion of the radial ground state in the inverted harmonic potential, as proposed in Refs.~\cite{Rom17,Pin18}, and show that this expansion preserves phase coherence. The experimental sequence is illustrated in Fig.~\ref{fig:Exp_protocal}(a). After preparing the atoms in the radial ground state, we switch on the RF field within $\sim 1\,\mu$s, inducing a rapid transverse quench from the single well onto the top of the inverted potential (Fig.~\ref{fig:Interference}(a)), with trapping frequency $\omega_I = 2\pi\times 0.9$~kHz set by the RF amplitude. Placed at the unstable fixed point, the cloud expands in both position and momentum: the initial interaction-modified transverse ground state is squeezed, its Wigner distribution elongating along one diagonal of phase-space while being compressed along the orthogonal direction, and both the spatial width $\sigma_x$ and the momentum width $\sigma_p$ grow. We refer to the resulting state as the \emph{expanded} state. For convenience we normalize position and momentum as 
\begin{equation} \nonumber
\begin{aligned}
    \widetilde{x} =  x/x_0 = x/[\hbar/2m\omega_\bot]^{1/2}  \\
    \widetilde{p} =  p/p_0 = p/[\hbar m\omega_\bot/2]^{1/2} 
\end{aligned}
\end{equation}
and drop the tilde over operators in the following.

That this expansion is coherent, the inflating state remains a single, phase-coherent quantum state rather than a stretched classical ensemble, can be demonstrated directly by matter-wave interference. By extending the duration of the double-well potential beyond the expansion stage, the cloud splits into two daughter clouds whose subsequent overlap produces an interference pattern. Figure~\ref{fig:Interference}(b) shows the resulting fringes integrated along the longitudinal direction after a total hold of $5.5$~ms, the full inflation on the IHO followed by roughly eight oscillation periods of the two daughter clouds in the double well. The observed fringe contrast $C = 0.24(0.04)$ and the peaked phase distribution in Fig.~\ref{fig:Interference}(c) demonstrate that phase coherence between the two copies survives far beyond the inflation duration and beyond the timescales accessed by the loop protocol below. 

\begin{figure}
\includegraphics[scale=0.29]{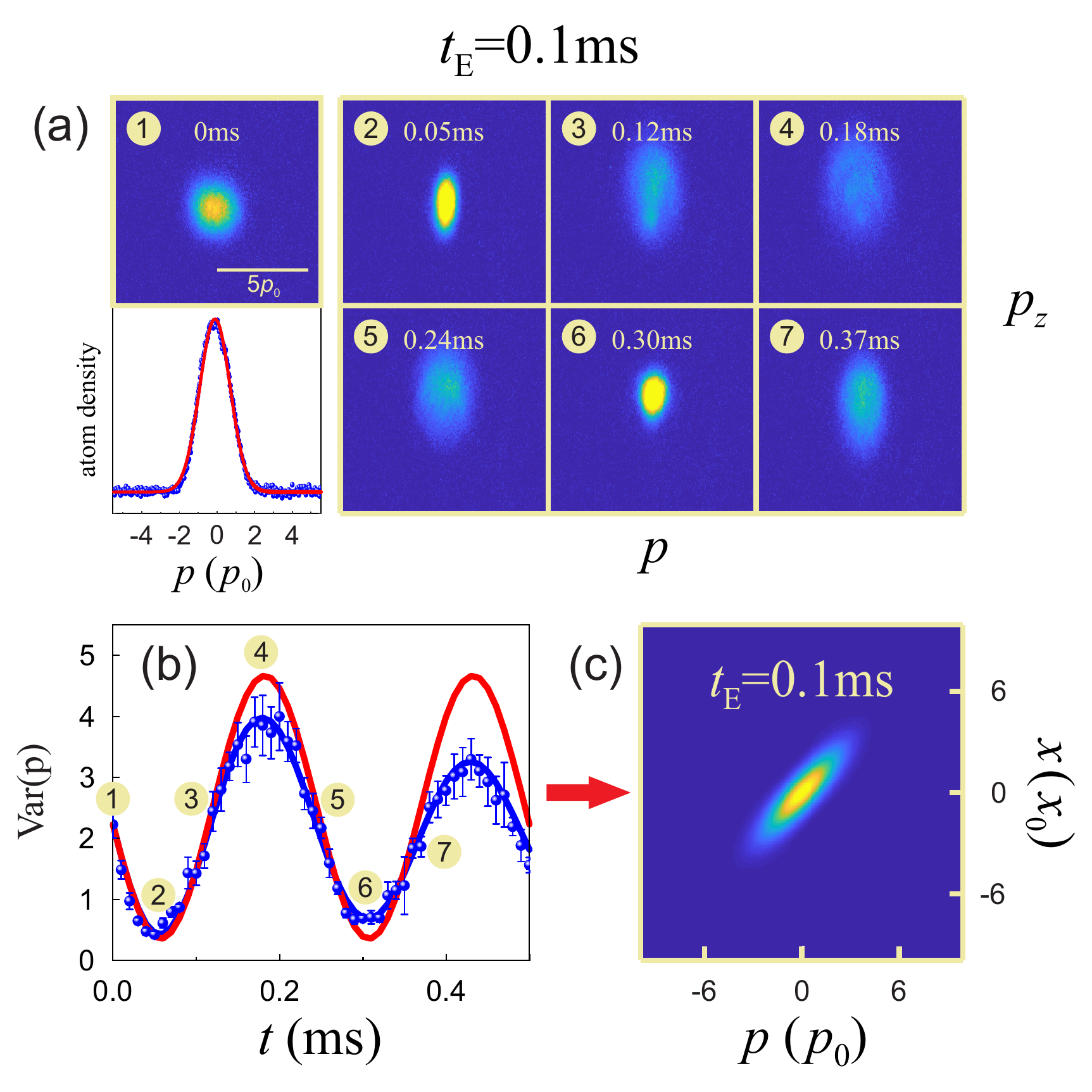}
\caption{\label{fig:Tomography}
(a) Momentum distributions at different tomography times for the expanded state with $t_E=0.1$~ms. Lower-left panel: after ToF the atomic density is integrated along $y$, and the momentum distribution along the expanded direction is fitted (red line) to extract its width and variance. (b) Evolution of the momentum variance, fitted with a damped-oscillation model (blue line); the red line is the corresponding undamped oscillation. The numbers mark the time points in panel (a). (c) Wigner distribution reconstructed from (b) by inverse-Radon transformation, for $t_E=0.1$~ms.}
\end{figure}

\vspace{5mm}\noindent\textbf{Phase-space tomography} \\
To fully characterize the expanded state we reconstruct its Wigner function by phase-space tomography. Limited imaging resolution prevents direct access to the position distribution; the momentum distribution, however, is measured accurately by absorption imaging after time of flight. We therefore quench the potential back to the initial harmonic trap by instantaneously switching off the RF currents. In the harmonic trap the Wigner distribution of the expanded state rotates in phase-space, and sampling the momentum distribution at different rotation angles allows the full distribution to be reconstructed by the inverse-Radon transformation~\cite{Leo97} (see Supplementary Information). The rotation frequency is set by the harmonic trapping frequency, $\sim 2\pi\times 2$~kHz, slightly modified by atomic interactions and the transverse motion of the cloud; we vary the tomography duration from $0$ to $0.5$~ms in steps of $0.01$~ms, covering one full rotation period. Taking $t_E=0.1$~ms as an example, Fig.~\ref{fig:Tomography}(a) shows the measured momentum distributions for the different holding times.

\begin{figure}[t]
\includegraphics[scale=0.29]{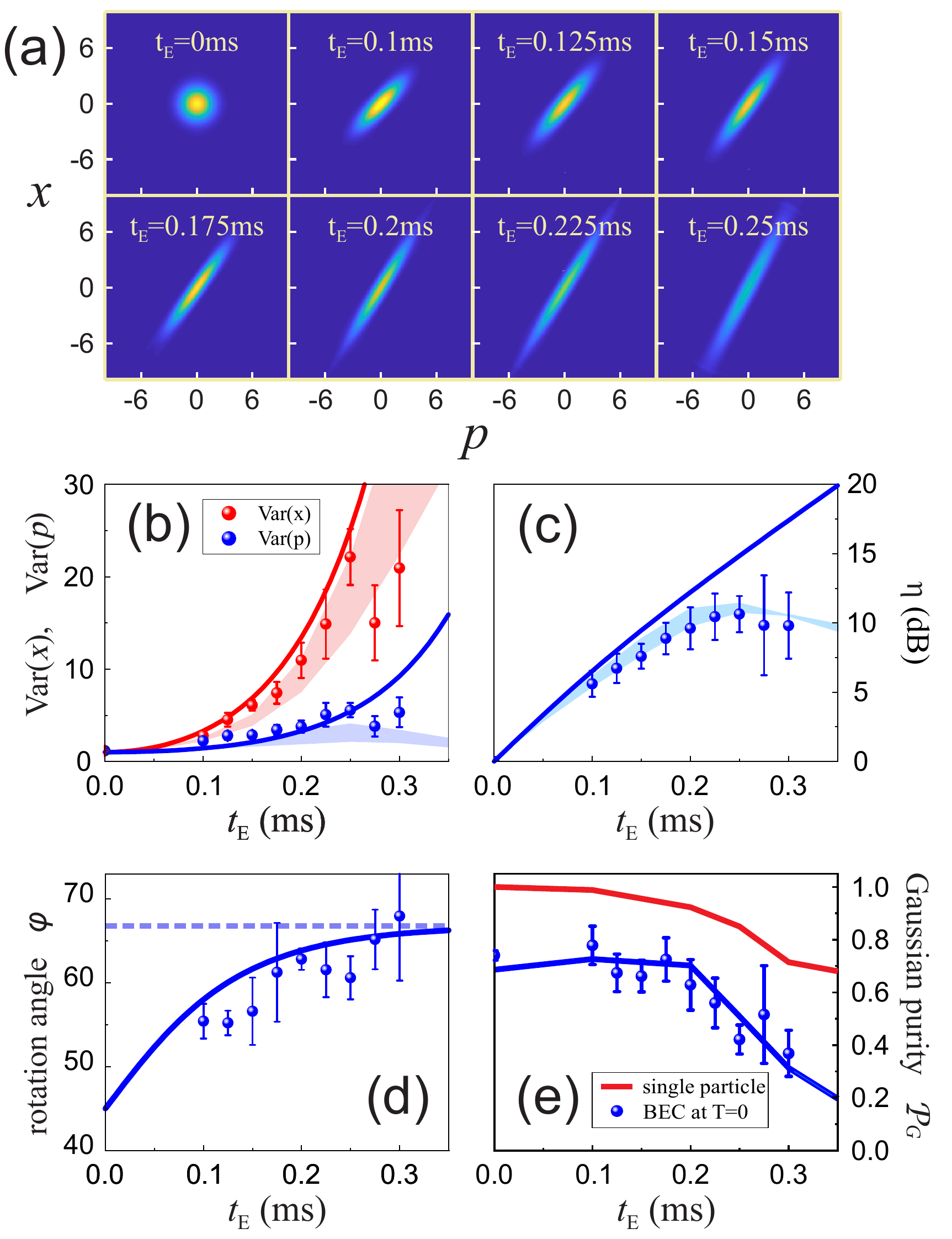}
\caption{\label{fig:Squeezing}
(a) Wigner distributions of the expanded states reconstructed by inverse-Radon transformation for varying expansion times.
(b) Momentum (blue) and position (red) variances of the expanded states versus expansion time.
(c) Squeeze factor versus expansion time.
(d) Orientation angle $\varphi$ of the major axis of the Wigner ellipse versus expansion time $t_E$. The asymptotic angle determined by $\omega_I/\omega_\bot$ is shown by the dashed line.
(e) Purity $\mathcal{P}_G$ of the reconstructed Gaussian state versus expansion time. The reduced $\mathcal{P}_G(0)\sim 0.75$ for the BEC results from interactions in the ground state; the apparent loss of purity with time, present even for a single particle, is the result of the anharmonicity of the implemented IHO.
In panels (b)–(d), solid lines denote the ideal single-particle IHO prediction; shaded bands denote the 2D-GPE simulation including experimental uncertainties. In panel (e), the red curve shows the apparent Gaussian purity for a single particle in the implemented anharmonic potential, while the blue line shows the 2D-GPE simulation for a BEC.}
\end{figure}

For the expansion we use the magnetically trappable state $|F=2,m_F=+2\rangle$. To suppress transfer into other Zeeman sublevels during the harmonic-to-inverted transition we control the phase of the RF field at the moment of the quench, keeping the population of undesired states below $10\%$; to mitigate the residual occupation of $|F=2,m_F=+1\rangle$ we extract the variance of the squeezed state from a Gaussian fit to the momentum distribution rather than from its raw second moment. The fitted width (lower-left panel of Fig.~\ref{fig:Tomography}(a)) yields the variance, normalized to $\mathrm{Var}_0(x)=\hbar/(2m\omega_\perp)$ and $\mathrm{Var}_0(p)=\hbar m\omega_\perp/2$. For each expansion time we record two full rotation periods (Fig.~\ref{fig:Tomography}(b)), with each data point averaged over 50--100 repetitions. A noticeable damping of the variance oscillation is visible in Fig.~\ref{fig:Tomography}(b); it originates in the tomography stage itself, from atomic interactions during the harmonic hold, and is unrelated to the expanded state prepared beforehand. We remove it by a damped-oscillation fit (blue line) and reconstruct the state from the undamped oscillation (red line; see Supplementary Information). The resulting Wigner distribution is shown in Fig.~\ref{fig:Tomography}(c), and the reconstructions for the full set of expansion times are summarized in Fig.~\ref{fig:Squeezing}(a).

The orientation of the reconstructed Wigner ellipse provides a parameter-free test of IHO dynamics. For ideal inverted-harmonic evolution from an isotropic coherent state, the major-axis angle $\varphi(t_E)$ is fixed solely by the frequency ratio $\omega_I/\omega_\perp$: it starts at $45^\circ$ at $t_E=0$ and approaches an asymptotic value set by the trap geometry. Figure~\ref{fig:Squeezing}(d) shows the measured angle together with the prediction of Eq.~(\ref{eq:squeezing_angle}), evaluated with the independently determined $\omega_I = 2\pi\times 0.9$~kHz and no adjustable parameters. The data follow the prediction across the whole range, reproducing both the initial $45^\circ$ symmetry of the single-transverse-mode, low-temperature initial state and the asymptotic plateau. Together with the variance growth shown in Fig.~\ref{fig:Squeezing}(b), this closes a self-consistency check: the same $\omega_I$ sets the potential geometry, the rate of phase-space stretching, and the orientation of the squeezed quadrature.

\vspace{5mm}\noindent\textbf{Sub-vacuum squeezing} \\
From the reconstructed Wigner distributions we extract the momentum and position variances, plotted in Fig.~\ref{fig:Squeezing}(b). Both grow exponentially with expansion time, as expected for hyperbolic phase-space flow. At longer expansion times ($t_E>0.25$~ms) both variances saturate and depart from the single-particle coherent-inflation prediction (red line)~\cite{Pin18}, with $\mathrm{Var}(x)$ falling consistently below the ideal expectation; we return to the origin of this deviation below.

We quantify the squeezing by the squeeze factor
\begin{equation} \label{eq:squeeze_factor}
\eta=-10\cdot \mathrm{log}_{10}\frac{\mathrm{Var}_{min}(p)}{\mathrm{Var}_0(p)},
\end{equation}
where $\mathrm{Var}_{min}(p)$ is the minimum momentum variance measured over the tomography cycle and $\mathrm{Var}_0(p)$ the variance of the initial state. Figure~\ref{fig:Squeezing}(c) shows $\eta(t_E)$; a maximum of $\eta=10.6(1.3)$~dB is reached at $t_E=0.25$~ms. Since $\mathrm{Var}_0(p)=\hbar m\omega_\perp/2$ is the vacuum variance of of the corresponding non-interacting harmonic-oscillator ground state, $\eta>0$~dB directly certifies sub-vacuum squeezing of the compressed quadrature: the corresponding fluctuations are pushed below the zero-point level. This squeezing below the harmonic-oscillator zero-point variance is a nonclassical signature of the amplified initial quantum fluctuations.

\vspace{5mm}\noindent\textbf{Purity and the role of interactions} \\

We quantify the departure from ideal single-mode Gaussian IHO evolution by the Gaussian purity $\mathcal{P}_G = 1/\sqrt{\det\Sigma}$, where $\Sigma$ is the covariance matrix in the normalized $(x,p)$ variables. For an ideal Gaussian state under a perfectly harmonic IHO, $\mathcal{P}_G$ is conserved. A reduction of $\mathcal{P}_G$, however, should be interpreted as an effective single-mode Gaussian purity loss: it can arise from anharmonic phase-space shearing, from coupling to transverse or longitudinal modes, or from genuine decoherence. The comparison with the GPE simulation shows that the observed reduction is dominated by deterministic interaction-induced mode mixing and anharmonicity, rather than environmental noise. Figure~\ref{fig:Squeezing}(e) shows $\mathcal{P_G}(t_E)$ from the same reconstructions as panels (a)--(d).

\begin{figure}
\includegraphics[scale=0.3]{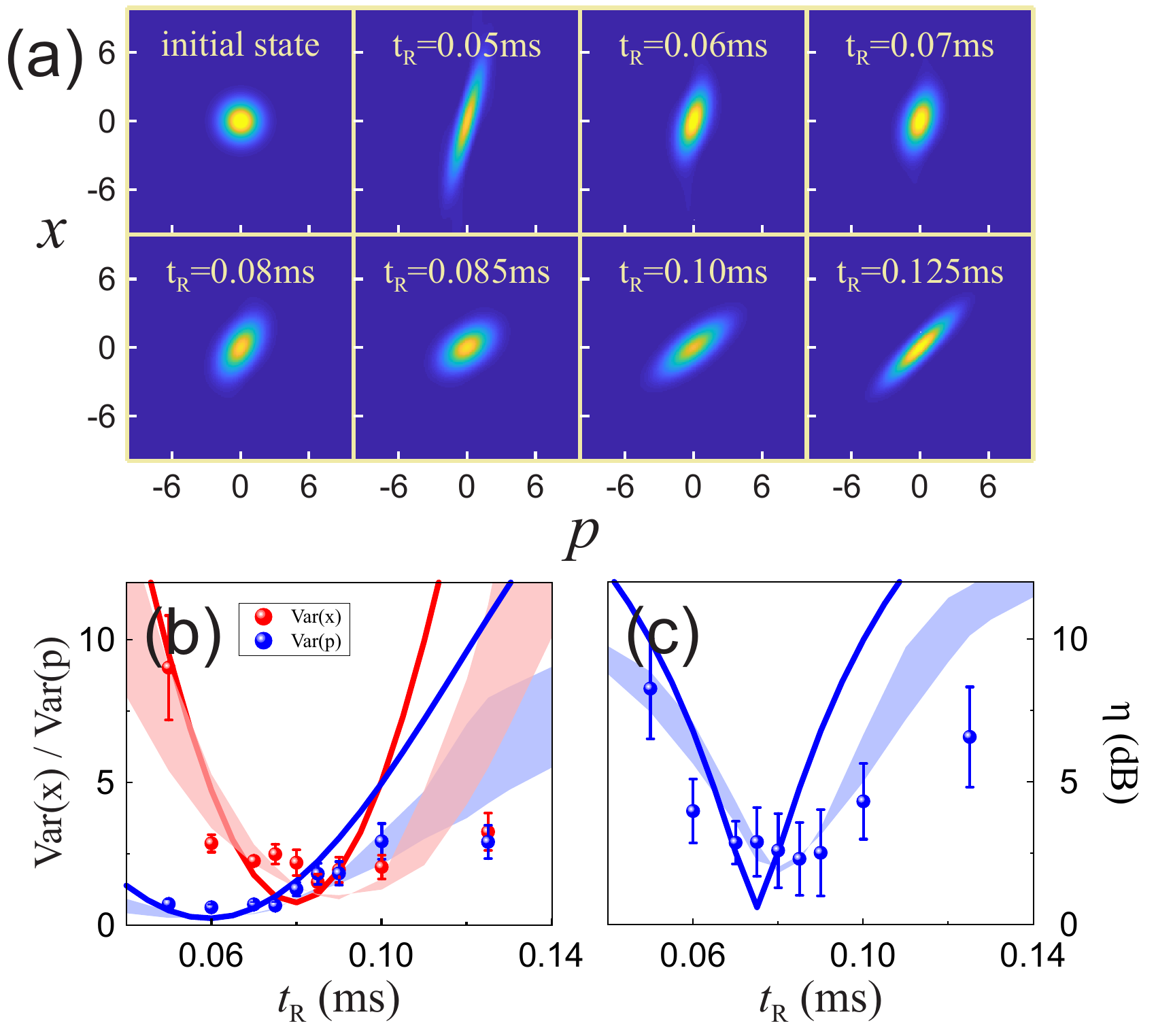}
\caption{\label{fig:Recover}
(a) Wigner distributions of the recovered states reconstructed by inverse-Radon transformation for varying rotation times; expansion and compression durations are both $0.15$~ms. (b) Momentum (blue) and position (red) variances of the recovered states versus rotation time. (c) Squeeze factor versus rotation time. For (b) and (c) the lines are the theoretical calculation and the shaded band is the GPE simulation including experimental errors.}
\end{figure}

The reduced initial Gaussian purity is a consequence of interactions in the ground state, while its subsequent decay reflects the combined effects of anharmonic phase-space shearing and interaction-induced mode mixing. In our system these are not imperfections but the physics of our experimental implementation: The initial value $\mathcal{P}_G(0)\approx 0.75$ is not set by thermal excitation, they are negligible for the radial ground state at $T\lesssim 40$~nK, where $\hbar\omega_\perp/k_B\approx 100$~nK \cite{Kruger2010}, but by the mean-field broadening of the interacting ground state. With a chemical potential $\mu_\perp$ just below $\hbar\omega_\perp$, the equivalent transverse single particle wave function is wider than the bare harmonic ground state, so a Gaussian fit to the momentum distribution registers the non-Gaussianity of the genuine many-body state as apparent mixedness. The expanded state thus inherits, from the outset, the interacting character of the condensate. We reproduce this offset, together with the squeezing and angle data, by a numerical simulation based on the 2D Gross--Pitaevskii equation (GPE) that incorporates the measured double-well potential and the atomic interactions (red and blue shaded bands in Figs.~\ref{fig:Squeezing} and~\ref{fig:Recover}).

The purity stays close to its initial value up to $t_E\approx 0.15$~ms and then falls to $\mathcal{P}_G\approx 0.4$ by $t_E=0.3$~ms. Crucially, this decay sets in at the same expansion time as the saturation of $\mathrm{Var}(p)$ in Fig.~\ref{fig:Squeezing}(b). The common origin is interaction-induced coupling between the principal squeezed mode and the remaining degrees of freedom --- not the coherent IHO dynamics itself --- which marks the onset of genuinely many-body behavior as the single-mode description begins to break down~\cite{Rod24,Cas24}. For $t_E\lesssim 0.15$~ms the protocol operates in the regime where the single-mode IHO description is accurate, and the sub-vacuum squeezing reported above amplifies genuine zero-point fluctuations rather than thermal noise; beyond it, the same interactions that broaden the initial state drive the crossover out of the single-mode picture. This controlled onset of many-body content is precisely what distinguishes the condensate platform from the single-particle, classical--thermal regime accessed with levitated nanoparticles~\cite{Tomassi2026}.

Several effects of the implementation limit the magnitude of the \emph{measured} squeezing, and are quantitatively captured by the  GPE. First, the inverted potential is harmonic only within $\sim 1\,\mu$m of the barrier top (see Supplementary Information); at longer expansion times the atoms sample the anharmonic shoulders of the barrier and the Gaussian inflation saturates. Second, during expansion and tomography the wide momentum distribution lets the fastest atoms escape the magnetic trap and transfers energy into the longitudinal direction, both of which reduce the measured $\mathrm{Var}(x)$ and $\mathrm{Var}(p)$ below the GPE prediction at large $t_E$; the interaction-induced damping of the tomography oscillation, corrected as described above, is a third, readout-stage effect. None of these alters the conclusions for $t_E\lesssim 0.15$~ms, where data, single-particle theory and GPE agree.

\vspace{5mm}\noindent\textbf{Time-reversal: the loop protocol} \\
To test the coherent reversibility of the inflation dynamics, we implement the loop protocol of Ref.~\cite{Wei21}: following the expansion stage, the squeezed state is allowed to rotate freely in the harmonic trap for a duration $t_R$, after which a second quench to the inverted potential for a compression time $t_C = t_E$ time-reverses the IHO evolution (Fig.~\ref{fig:Exp_protocal}(b)). For unitary evolution in a perfect IHO the final state would coincide with the initial state, with squeezing factor $\eta = 0$. Imperfections like deviations from an ideal harmonic oscillator potential, the interaction-induced single-mode dephasing as well as decoherence mechanisms like magnetic field noise, collisions with background particles will broaden and distort the final Wigner distribution.

We fix $t_C = t_E = 0.15$~ms and scan the rotation duration $t_R$. The reconstructed Wigner distributions (Fig.~\ref{fig:Recover}(a)) approach the initial state most closely at $t_R = 0.085$~ms, where the residual squeezing factor reaches its minimum (Fig.~\ref{fig:Recover}(c)). Comparing to the single-particle CI prediction (solid lines), we observe a shift of the optimal rotation time and a residual $\eta_{\min} > 0$. Both effects are captured by the 2D-GPE simulation (shaded bands), which incorporates the measured double-well potential and the atomic interactions. The agreement between data and GPE across the full loop confirms that the dominant source of imperfect refocusing is interaction-induced mode coupling, consistent with the purity analysis of the expansion stage, while environmental noise contributes negligibly. The loop protocol thus demonstrates that the inflation is reversible to within a known and quantitatively understood mechanism.

\vspace{5mm}\noindent\textbf{Discussion and Conclusion } \\
We have realized the quantum regime of coherent inflation with a Bose--Einstein condensate, using fast radio-frequency control on an AtomChip to prepare $\sim 10^4$ atoms at the unstable fixed point of an inverted harmonic oscillator and to characterize the resulting evolution by full Wigner tomography at every stage. The exponential growth of both quadrature variances, the parameter-free rotation of the Wigner ellipse toward the asymptotic angle set by $\omega_I/\omega_\perp$, the sub-vacuum squeezing of $10.6(1.3)$~dB, and the substantial restoration of the initial state after time-reversing the IHO evolution in a loop protocol together demonstrate that the inflation dynamics is genuinely quantum, coherently reversible, and quantitatively reproduced by Gross–Pitaevskii simulations within the regime of weak interaction-induced single-mode dephasing. The purity decay extracted directly from the reconstructed covariance matrix identifies the same interaction-driven mode coupling as the common origin of variance saturation and effective single-mode decoherence, isolating the regime ($t_E \lesssim 0.15$~ms) where the single-mode IHO description is accurate. In contrast to the classical--thermal regime accessed with levitated nanoparticles~\cite{Tomassi2026}, where the observed phase-space stretching applies equally to classical Gaussian distributions, the experiment reported here demonstrates the amplification of zero-point fluctuations themselves in a many-body quantum system.

The platform opens several directions. The matter-wave interference between the daughter clouds, sustained over many oscillation periods, indicates that the inverted potential can be operated in regimes far beyond the single-mode picture, where the dynamics couples to longitudinal degrees of freedom and produces phonon-like excitations along the elongated trap axis. Controlling the longitudinal trapping potential~\cite{Tajik:19} and imaging the local relative phase along this z-axis gives access to mode mixing and quasiparticle production, with a direct analogy to the squeezing of inflationary perturbations on a de~Sitter background~\cite{Albrecht1994} and to reheating dynamics in quantum field theory~\cite{Kofman1994}. Combined with the loop protocol, the  exponential amplification of weak displacements provides a route to quantum-enhanced force sensing with time-reversal-based certification of coherence~\cite{Geraci2015,Hebestreit2018}. Finally, the demonstrated control over a pure, low-noise IHO evolution in a many-body system provides a starting point for studying beyond-mean-field corrections to coherent inflation~\cite{Rod24,Cas24}, where the single-mode description begins to break down and the genuinely many-body content of the dynamics is expected to emerge.
 
\vspace{5mm}\noindent\textbf{Acknowledgments} \\
We thank T. Schweigler, B. Rauer, J. Sabino for help with the experimental setup and M. Serbyn for helpful discussions. 
This research was supported by the CoE: \textit{Quantum Science Austria} (QuantA), the European Research Council: ERC-AdG “Emergence in Quantum Physics” (EmQ) under Grant Agreement No. 101097858,  the DFG / FWF Collaborative Research Center SFB 1225: ISOQUANT (FWF: I6949) and by the Austrian Science Fund (FWF) Grant No. I6276 (QuFT-Lab).

\bibliographystyle{apsrev4-2}
\bibliography{Coherent_Inflation_citation_merged}

\newpage

\setcounter{figure}{0}
\setcounter{equation}{0}
\renewcommand{\theequation}{S\arabic{equation}}
\renewcommand{\thefigure}{S\arabic{figure}}
\renewcommand{\thetable}{S\arabic{table}}

\section*{supplementary information}

\subsection{Experimental set-up and radio-frequency dressed potential}

A one-dimensional quasicondensate of $^{87}\mathrm{Rb}$ is prepared using standard magneto-optical trapping and laser cooling techniques, followed by radio-frequency (RF) knife evaporative cooling to reach the quantum degeneracy. The atoms are tightly confined in transverse directions, with a transverse trapping frequency of $\omega_\perp=2\pi\times 2.1\mathrm{kHz}$. In contrast, weak harmonic confinement along the longitudinal direction is provided by an additional pair of U-shaped wires on the AtomChip, resulting in a longitudinal trapping frequency of $\omega_l=2\pi\times 10\mathrm{Hz}$. After evaporative cooling, the quasicondensate contains approximately $N\approx 10^4$ atoms at a peak linear density of $n_{1d} \simeq 100$ atoms/µm and a temperature in the range $T=20-40\mathrm{nK}$.  With both $\mu_\perp < \hbar \omega_\perp$ and $k_B T < \hbar \omega_\perp$ the transverse wavefunction of the BEC is well described by an interaction-broadened Gaussian with negligible transverse excitations \cite{Kruger2010}. 

By applying RF currents through a pair of auxiliary wires adjacent to the main wire on the chip (Fig.~\ref{fig:MagneticPotential}a), the Zeeman shifts of the magnetic sublevels acquire a spatial dependence. The resulting RF-dressed potential can be locally engineered by adjusting the amplitude, imbalance, and relative phase of the two RF currents. For balanced RF currents with a $\pi$ relative phase, the trap is stretched along the horizontal (x) direction while the harmonic confinement in the orthogonal transverse (y) direction remains unchanged. When the peak-to-peak amplitude of RF exceeds a critical value, the initially single-well potential is continuously deformed into a double-well configuration.

\subsection{Simulation of dressed magnetic trap and the distinction between double-well and inverted harmonic potentials}
\begin{figure}[t]
\includegraphics[width=\columnwidth]{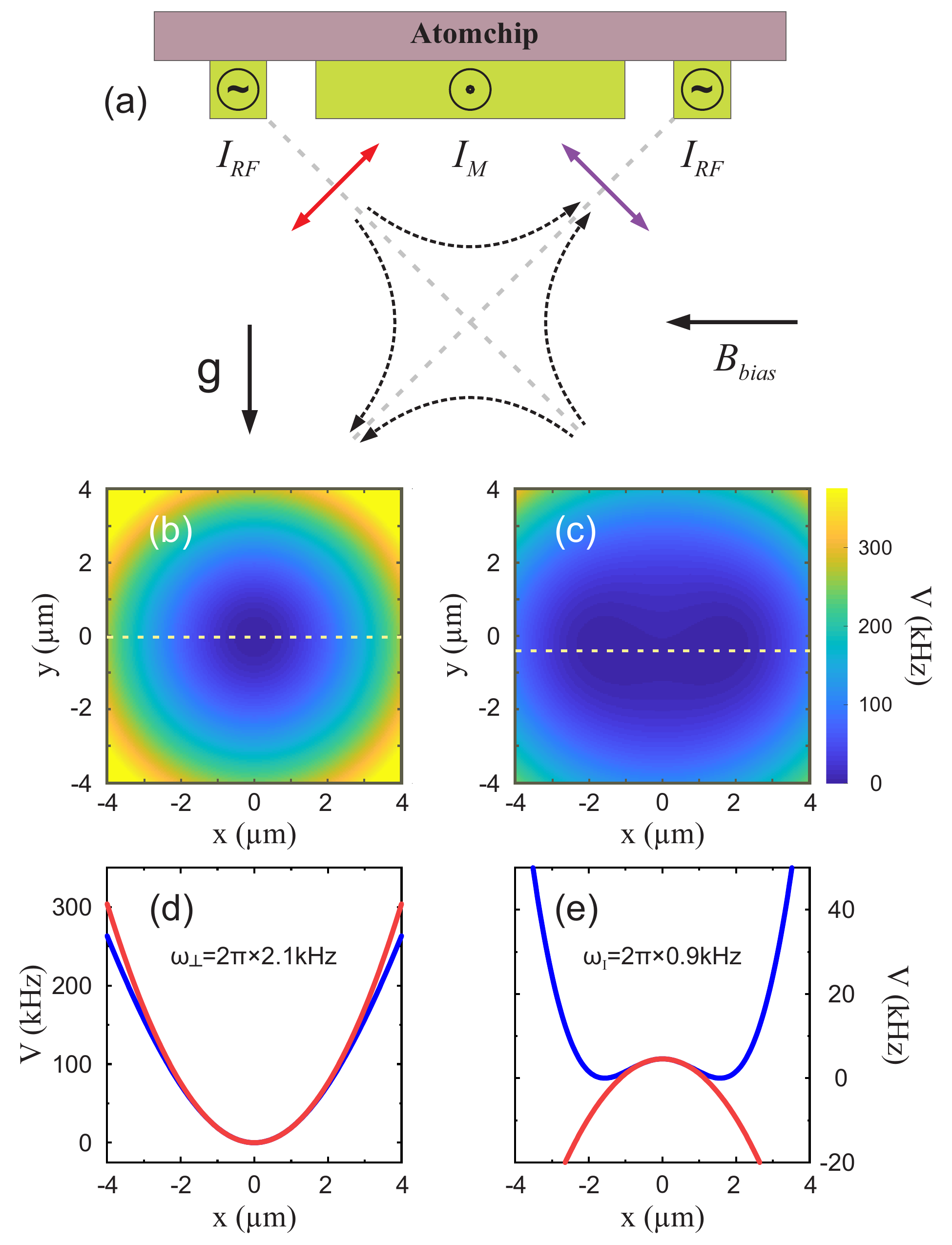}
\caption{\label{fig:MagneticPotential} 
(a) Schematic of the experimental setup at the AtomChip, showing the cross-section of the wires employed for the magnetic trap and the RF field.
Simulated magnetic potentials for the undressed single-well (b) and double-well (c) configurations. The trapping potential along the x direction near the energy minimum (d) and the barrier (e) are shown by the blue curves, while the red curves represent the corresponding approximations by a quadratic harmonic oscillator potential. }
\end{figure}

In this experiment, the current in the main wire on AtomChip is set to 800 mA, and a bias magnetic field of 14.5 Gauss is applied below the chip (Fig.~\ref{fig:MagneticPotential}(a)).  
For the double-well, an additional radio-frequency current with a peak-to-peak amplitude $I_{RF}=50$mA is applied. To compare with the harmonic potential, we simulate the spatial distribution of the magnetic field in Fig.~\ref{fig:MagneticPotential}.
 Figure~\ref{fig:MagneticPotential} (b) shows the cross-sectional magnetic potential of the undressed single well. By fitting the potential near the minimum with a parabolic function (Fig.~\ref {fig:MagneticPotential} d), we extract the initial trapping frequency, $\omega_\perp=2\pi\times 2.1\mathrm{kHz}$, which is consistent with the trapping frequency measured by collective excitation. Fig.~\ref{fig:MagneticPotential} (c) displays the double-well potential with a central barrier. In Fig.~\ref{fig:MagneticPotential} (e), the red curve shows a parabolic fit to the center region, from which the trapping frequency of the inverted harmonic potential is obtained as $\omega_I=2\pi\times 0.9\mathrm{kHz}$.

In Fig~\ref{fig:MagneticPotential}, we compare the inverted harmonic trap and the double-well potential. We find that within the region $[-0.95,+0.95]\: \mu$m, the relative difference is smaller than $15\%$, indicating that this barrier region can be well approximated by an inverted harmonic potential.

\subsection{Inverse Radon transformation and reconstruct Wigner distribution}

In the experiment, after the time-of-flight, by measuring the width of atomic cloud along x direction with different tomography time (Fig.~\ref{fig:Tomography}(b)), we have obtained the momentum distribution $pr(p, \theta)=\frac{1}{\sqrt{2\pi \mathrm{Var}(p_\theta)}}exp(-\frac{p^2}{2\mathrm{Var}(p_\theta)})$ for various rotation angles: $\theta=\omega_\bot t=0:\pi/25:2\pi$. This distribution can be expressed as the integral of the rotated Wigner function over the position direction: 
\begin{equation} \label{eq:projection}
\begin{aligned}
pr(p,\theta)=\int_{-\infty}^{+\infty} W(x\cos\theta-p\sin\theta, x\sin\theta+p\cos\theta) \,dx.
\end{aligned}
\end{equation}
Then, by leveraging these complementary features of the momentum distribution, we can deduce the complete information  of the squeezed state using the inverse-Radon transformation \cite{Leo97}:
\begin{equation} \label{eq:inverse_Radon}
\begin{aligned}
W(x,p)=\frac{\mathcal{PV}}{2\pi^2}\int^{\pi}_0\int^{+\infty}_{-\infty}\frac{pr(p_\theta,\theta)dp_{\theta}d\theta}{(-x\sin\theta+p\cos\theta-p_{\theta})^2},
\end{aligned}
\end{equation}
here, $\mathcal{PV}$ denotes the Cauchy principal value, introduced due to the singularity at the zero point. Note that $p_{\theta}$ and $p$ are distinct: $p_{\theta}$ corresponds to the momentum at tomography angle $\theta$, while $p$ represents the momentum of the expanded state we wish to reconstruct (i.e., momentum with $\theta=0$).

\subsection{2D GPE simulation}
In this experiment, owing to the pronounced separation of trapping frequencies between the longitudinal and transverse directions, with a transverse confinement of 2.1 kHz and a much weaker longitudinal confinement of 10 Hz, the characteristic dynamical time scales along these axes differ by more than two orders of magnitude. As a result, in the short-time limit, the dynamics of the system in the longitudinal and transverse directions are effectively decoupled. For simplicity, the longitudinal degree of freedom is treated as homogeneous, and the transverse dynamics can be faithfully described by an effective two-dimensional Gross-Pitaevskii equation (2D GPE). This approach enables a quantitative description of coherent inflation and the loop protocol induced by rapid modifications of the transverse trapping potential.

The effective 2D GPE governing the transverse wavefunction $\psi_\perp(x, y , t)$ reads:

\begin{equation} \label{eq:2DGPE}
\begin{aligned}
i\hbar \frac{\partial \psi_\perp(x,y,t)}{\partial t}
=& \left[ -\frac{\hbar^2}{2m}\nabla^2 + V(x,y,t) \right. \\
& \left. + g_{3\mathrm{d}}\, n_{1\mathrm{d}} \, \left|\psi_\perp(x,y,t)\right|^2 \right] \psi_\perp(x,y,t),
\end{aligned}
\end{equation}
where V is the time-dependent transverse potential and $g_{3\mathrm{d}}=4\pi\hbar^2a_s/m$ is the three-dimensional interaction strength. Here, $n_{1\mathrm{d}}$ represents the typical linear atomic density at the center of the trap, taken to be $n_{1\mathrm{d}}=100 \mu \mathrm{m}^{-1}$, and the transverse wavefunction is normalized to unity.

At longer evolution times, however, excitations are expected to develop along the longitudinal (z) direction, giving rise to quasiparticles such as phonons and solitons, which invalidate the decoupling assumption. Accessing these dynamics requires imaging along the vertical direction, enabling the investigation of both relative phase and density. Characterizing these coupled longitudinal–transverse excitation modes constitutes an important direction for future investigations.

\subsection{Coherent Inflation in an inverted harmonic trap}
The Hamiltonian of particles at the top of an inverted harmonic trap is:
\begin{equation} \label{eq:Hamiltonian_Inv}
\begin{aligned}
H=\frac{p^2}{2m}-\frac{m\omega_I^2x^2}{2}=\frac{p^2}{2m}+\frac{m(i\omega_I)^2x^2}{2}.
\end{aligned}
\end{equation}
The system can be treated as a harmonic oscillator with a complex frequency: $i\omega_I$.
Analogous to a harmonic oscillator, the equations of motion for the inverted potential
take the form:
\begin{equation} \label{eq:EoM1}
\left\{
\begin{aligned}
x(t)&=x_0\cos(i\omega_It)+\frac{p_0}{im\omega_I}\sin(i\omega_It)\\
p(t)&=p_0\cos(i\omega_It)-im\omega_Ix_0\sin(i\omega_It).
\end{aligned}
\right.
\end{equation}
By using hyperbolic functions, we can simplify the above equations to:
\begin{equation} \label{eq:EoM2}
\left\{
\begin{aligned}
x(t)&=x_0\cosh(\omega_It)+\frac{p_0}{m\omega_I}\sinh(\omega_It)\\
p(t)&=p_0\cosh(\omega_It)+m\omega_Ix_0\sinh(\omega_It).
\end{aligned}
\right.
\end{equation}
Starting from a coherent state characterized by initial variances $v_{x0}$ and $v_{p0}$ in position and momentum, respectively, satisfying $v_{x0}=v_{p0}/m^2\omega_\bot^2$, where $\omega_\bot$ denotes the trapping frequency of the initial harmonic trap. We can further obtain the evolution of variances of a Gaussian squeezed state:
\begin{equation} \label{eq:Evolution_Var}
\left\{
\begin{aligned}
v_x(t)&=v_{x0}\cosh^2(\omega_It)+\frac{v_{p0}}{m^2\omega_I^2}\sinh^2(\omega_It)\\
v_p(t)&=v_{p0}\cosh^2(\omega_It)+m^2\omega_I^2v_{x0}\sinh^2(\omega_It).
\end{aligned}
\right.
\end{equation}
In the main text, the theoretical curves of $\mathrm{Var}(x)$ and $\mathrm{Var}(p)$ in Fig.~\ref{fig:Squeezing} (b), as well as the corresponding squeeze factor $\eta$ in Fig.~\ref{fig:Squeezing} (c) are derived from Eq.~\ref{eq:EoM2} and Eq.~\ref{eq:Evolution_Var}.

For simplicity, both axes are presented in dimensionless units, with position and momentum normalized by $x_0=(\hbar/2m\omega_\bot)^{1/2}$ and $p_0=(\hbar m\omega_\bot/2)^{1/2}$, respectively. Under this normalization, then the angle between the major axis of the Wigner elliptical distribution and the p-axis will vary with time:

\begin{equation} \label{eq:squeezing_angle}
\begin{aligned}
\varphi (t)=\frac{\pi}{2}-\frac{1}{2}\arctan \left( \frac{2\sqrt{v_x(t)v_p(t)-1}}{v_x(t)-v_p(t)} \right).
\end{aligned}
\end{equation}

At long time scales, both variances exhibit exponential growth:
\begin{equation} \label{eq:Evolution_Var2}
\left\{
\begin{aligned}
v_x(t)\approx(v_{x0}+\frac{\omega_\bot^2}{\omega_I^2}v_{p0})\frac{e^{2\omega_It}}{4}\\
v_p(t)\approx(v_{p0}+\frac{\omega_I^2}{\omega_\bot^2}v_{x0})\frac{e^{2\omega_It}}{4}.
\end{aligned}
\right.
\end{equation}

\begin{figure}
    \includegraphics[width=\columnwidth]{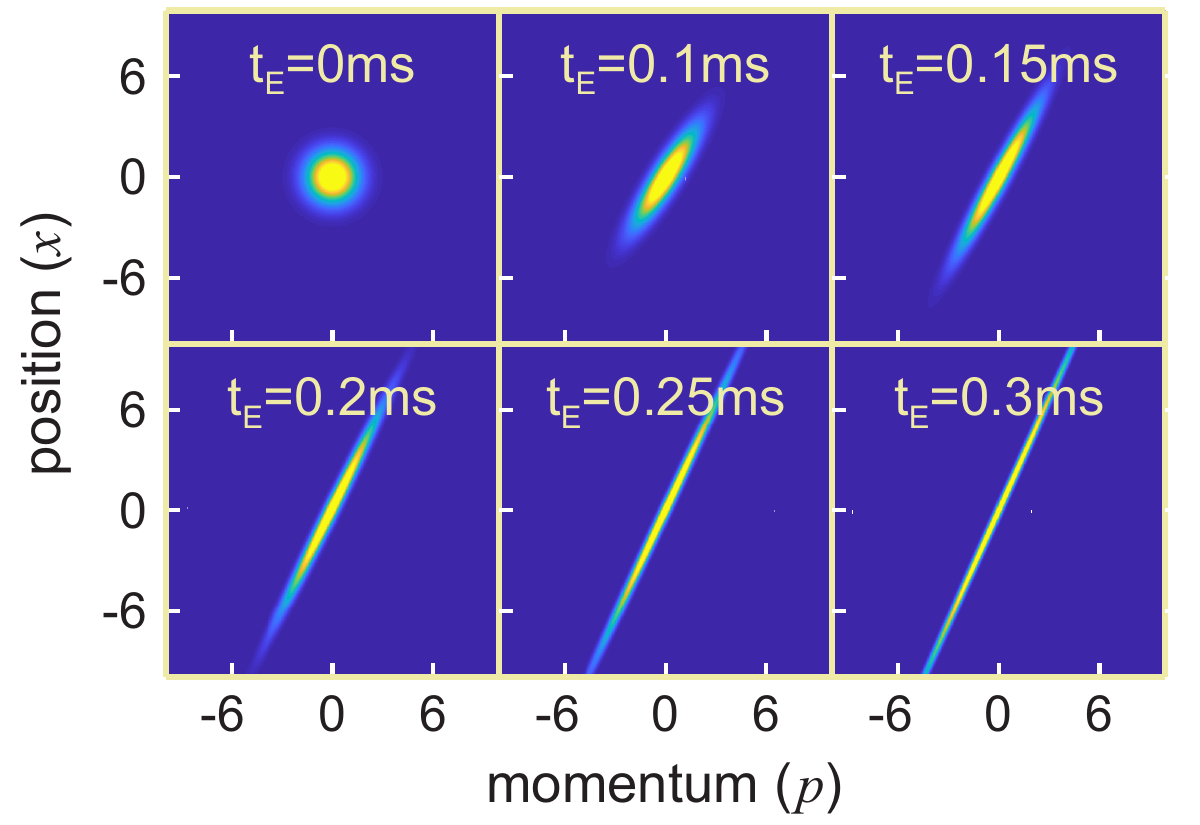}
    \caption{\label{fig:Wigner} 
    The evolution of the Wigner function for a squeezed state in an inverted harmonic trap.}
\end{figure}

For an initial state that is the ground state of the $\omega_\bot$ trap, with $v_{x0}=v_{p0}$, the squeezing angle reaches a plateau in the long-time limit, determined by  $\omega_I/\omega_\bot$:
\begin{equation} 
\begin{aligned}\label{eq:Longtime_squeezeAngle}
\varphi_{t\rightarrow\infty}=\arctan \left( \frac{\omega_\bot}{\omega_I} \right).
\end{aligned}
\end{equation}

Figure~\ref{fig:Wigner} illustrates the Wigner distribution for the expanded states with different expansion times in phase-space. 

\subsection{Damped-oscillation fit of Var(p)}
In Fig.~\ref{fig:Tomography} (b) of the main text the measured $\mathrm{Var}(p)$ during the tomography stage exhibits a damped oscillation. This damping arises from the coupling between atomic interaction and the collective motion of the atomic cloud along the other transverse (y) direction. Numerical simulation based on 2D-GPE reveals that both the oscillation amplitude and its offset decay over time, with approximately equal decay coefficients.

Based on the above facts, we model a damped oscillation of $\mathrm{Var}(p)$ using the following equation:
\begin{equation} \label{eq:Damped_Oscillation}
\begin{aligned}
\mathrm{Var}(p_t)=[A\cdot \cos(\omega t+\phi)+B]\cdot e^{(-t/\tau)}+C,
\end{aligned}
\end{equation}
which exhibits excellent agreement with the experimental data (blue line in Fig.~\ref{fig:Tomography} (b)). 

The damping in \ref{eq:Damped_Oscillation} originates from the tomography process, in order to reconstruct the state before tomography, we employ the non-damped oscillation model by taking out the exponential damping factor (red line in Fig.~\ref{fig:Tomography} (b)):
\begin{equation} \label{eq:Non_Damped_Oscillation}
\begin{aligned}
\widetilde{\mathrm{Var}}(p_t)= A \cdot \cos{(\omega t+\phi)}+B+C.
\end{aligned}
\end{equation}

\end{document}